\documentclass[preprint2,twoside]{hwo}

\usepackage{lipsum}

\input{hwo.h}

\begin{document}

\title{\textbf{\LARGE Keck Observatory as an HWO Testbed:\\ 
validating wavefront sensing and control schemes on a large segmented aperture in parallel with high-contrast science}}
\author {\textbf{\large Ma\"issa Salama,$^{1}$ Rebecca Jensen-Clem,$^1$ Mahawa Ciss\'e,$^2$ J. Kent Wallace,$^3$ Mitchell Troy,$^3$ Laurent Pueyo,$^4$ Charlotte Guthery,$^2$ Antonin Bouchez,$^2$ Vincent Chambouleyron$^5$}}
\affil{$^1$\small\it University of California, Santa Cruz, Santa Cruz, California, USA}
\affil{$^2$\small\it W. M. Keck Observatory, Hawaii, U.S.A.}
\affil{$^3$\small\it Jet Propulsion Laboratory, California Institute of Technology, U.S.A.}
\affil{$^4$\small\it Space Telescope Science Institute, Maryland, U.S.A.}
\affil{$^5$\small\it Laboratoire d'Astrophysique de Marseille, France}



\begin{abstract}
Exoplanet direct imaging allows us to directly probe and characterize an exoplanet’s atmosphere, searching for signs of life in its atmospheric signatures. Directly imaging an Earth-like planet around a Sun-like star requires reaching 10$^{-10}$ contrast levels and will be the goal of the Habitable Worlds Observatory (HWO). A key technical barrier to reaching such deep contrasts is maintaining wavefront stability on the order of tens of picometers, in particular in the presence of a segmented primary mirror. Keck Observatory is the only facility with all of the hardware components necessary for validating HWO segment phasing strategies: a large segmented primary mirror, capacitive edge sensors, deformable mirror, Zernike wavefront sensor (ZWFS), and high contrast science instruments. Taking advantage of these parallels, we are using Keck as a testbed for developing and validating HWO wavefront sensing and control loop strategies, as well as demonstrating the full system-level segment control architecture for HWO, using existing infrastructure. Recently, we set the stage for this work by using the ZWFS installed on the Keck II telescope to sense and correct the primary mirror segment pistons in closed-loop in parallel with science observations. This resulted in improved Strehl ratios on the NIRC2 science camera \citep{Salama24}. We now aim to directly address concerns related to control authority, actuator offload, and loop stability – tasks which require Keck's existing infrastructure, but which do not require picometer wavefront stability. Moreover, successful comparisons of observed and predicted performances will validate, on a real operating observatory, the HWO error budget methodology and in particular its approach to nested loops operating at multiple timescales.
  \\
\end{abstract}

\vspace{2cm}

\section{Introduction}

The key challenge for HWO is achieving the 10$^{-10}$ contrast at separations $\leq$0.1~arcseconds that are required for habitable zone (HZ) rocky planet imaging around G-type stars. Over the last several decades, a range of studies have linked this contrast requirement to a primary mirror wavefront stability requirement. For example, \cite{Trauger07} found that mid-scale primary mirror figure errors must be less than 14~pm RMS over 90 minutes. More recently, \cite{Nemati20} found that to detect exo-Earths using a coronagraph on a segmented aperture telescope requires segment-to-segment co-phasing better than 2~pm RMS. \cite{Pueyo22} point out that meeting such challenging requirements over all timescales is only necessary in the traditional ``set-and-forget" paradigm of segment co-phasing that was used for JWST. These co-phasing requirements can be relaxed if we consider methods of continuous, closed-loop wavefront control that take advantage of the broad suite of deformable mirrors (DMs) and wavefront sensors (WFSs) – in addition to the primary mirror segments themselves – that are available for maintaining high contrast throughout a science exposure. Ground-based high contrast imaging systems like Keck/NIRC2, the Gemini Planet Imager (GPI; \citealt{Macintosh18}) and Spectro-Polarimetric High-Contrast Exoplanet REsearch (SPHERE; \citealt{Beuzit19}) have long used wavefront sensing and control (WFS\&C) systems not only to correct for millisecond-timescale atmospheric variations, but also to correct for tens of minute timescale instrument and telescope-induced aberrations (albeit at different RMS wavefront error scales than those expected for HWO).

State-of-the-art high contrast testbeds such as NASA's High Contrast Imaging Testbed (HCIT; \citealt{Seo19}) and STScI's High-contrast imager for Complex Aperture Telescope (HiCAT; \citealt{Soummer24}) do not yet include subsystems for simulating the effects of realistic segmented primary mirrors. HiCAT has preliminarily addressed this need using a segmented DM, but the segments are flat, cannot be warped, and there is no clear path for implementing capacitive edge sensors in a way that would be analogous to a full-scale system. However, future laboratory-scale segmented apertures could not capture the full spatio-temporal dynamics of a many-meter sized, full-scale segmented mirror.

\subsection{Relevant HWO Technology Gaps}

Both \cite{Nemati20} and The Ultra-Stable Telescope Research and Analysis (ULTRA; \citealt{ULTRA}) Phase I report have identified that mid-spatial frequencies (especially segment-level piston-tip-tilt) will be most damaging to contrast stability in the coronagraphic data. To maintain contrast in the presence of segment primary wavefront drifts over a broad range of temporal scales, \cite{Coyle22} propose a combination of sensing and control with the science instrument and telescope metrology. \cite{Potier21} also show that an adaptive optics (AO)-like approach to correcting segment drifts and vibrations at $\sim$10~Hz can be successful for an HWO architecture, provided that the natural guide star is sufficiently bright and that segment motions show strong correlations. On the other hand, under the more conservative assumption of independent segments, the photon content of wavefront sensing images for a single segment becomes sufficiently limited that an instrument-based control loop (e.g. a Zernike WFS controlling a DM) can only operate with a DM update cadence on the order of minutes \citep{Potier22,Sahoo22}. Under this scenario, discussed in \cite{Coyle22}, telescope metrology that does not rely on the finite flux of a natural guide star is needed to address faster timescales. Experiments with capacitive edge sensors coupled with piezo-actuators have demonstrated that such an in-situ sensing and control architecture does yield a segment-to-segment $\sim$picometer phasing residual \citep{Cromey22} over timescales shorter than minutes, which subsequently drift due to electronics noise. While this noise is being mitigated \citep{Cromey22}, an architecture combining instrument wavefront maintenance over a long timescale and edge sensor wavefront maintenance over a short timescale seems the most promising for HWO.

\subsection{Keck Observatory as an HWO testbed}

\begin{figure}[ht!]
\begin{center}
\includegraphics[width=0.48\textwidth]{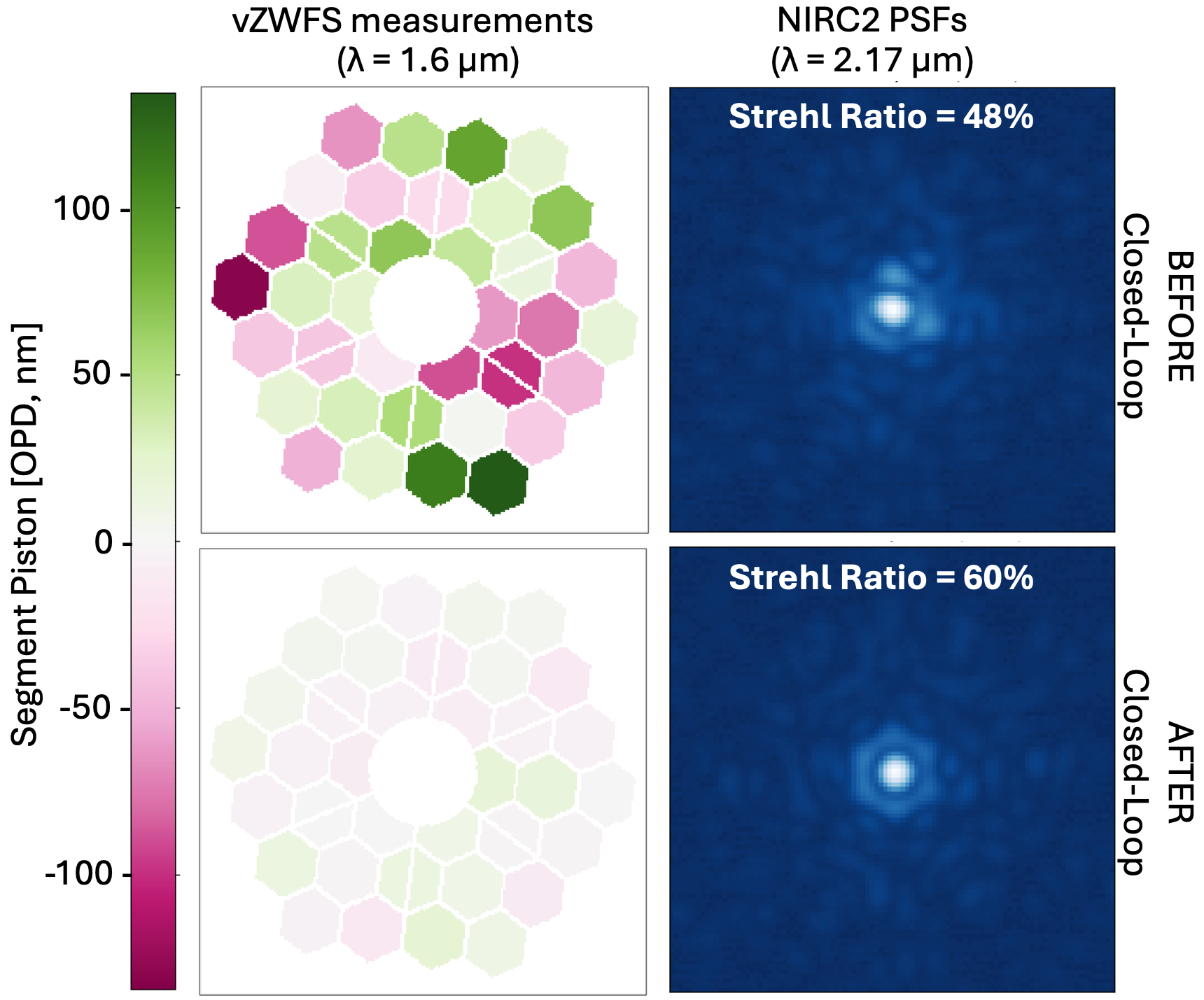}
\caption{\small Demonstration of using the ZWFS installed on the Keck II telescope to sense the primary mirror segment pistons and control them with the segment actuators in closed-loop, improving the Strehl ratios on the NIRC2 science camera. \textit{Left:} ZWFS segment piston measurements before (top) and after (bottom) the ZWFS closed-loop. \textit{Right:} NIRC2 PSF images before (top) and after (bottom) the ZWFS closed-loop. These measurements correspond to the 2025-06-17 Run 1 reported in \cite{Salama25}. Many such closed-loop runs have now been conducted \citep{Salama24,Salama24b,Salama25}.
\label{fig:ZWFS_onsky}
}
\end{center}
\end{figure}

The Keck observatory is the only facility in the world already equipped with the hardware to carry out such a wavefront maintenance strategy. We began this work in \cite{Salama24}, where we demonstrated the first closed-loop control of a segmented aperture telescope using a Zernike wavefront sensor (ZWFS) – the baseline wavefront sensor used in the LUVOIR concept study. Figure \ref{fig:ZWFS_onsky} illustrates our ability to improve the Strehl ratio on the NIRC2 science camera when we sense primary mirror segment co-phasing errors with the ZWFS and control them using the primary mirror segment actuators. 

\begin{figure*}[ht!]
\begin{center}
\begin{tabular}{c}
\includegraphics[width=0.95\textwidth]{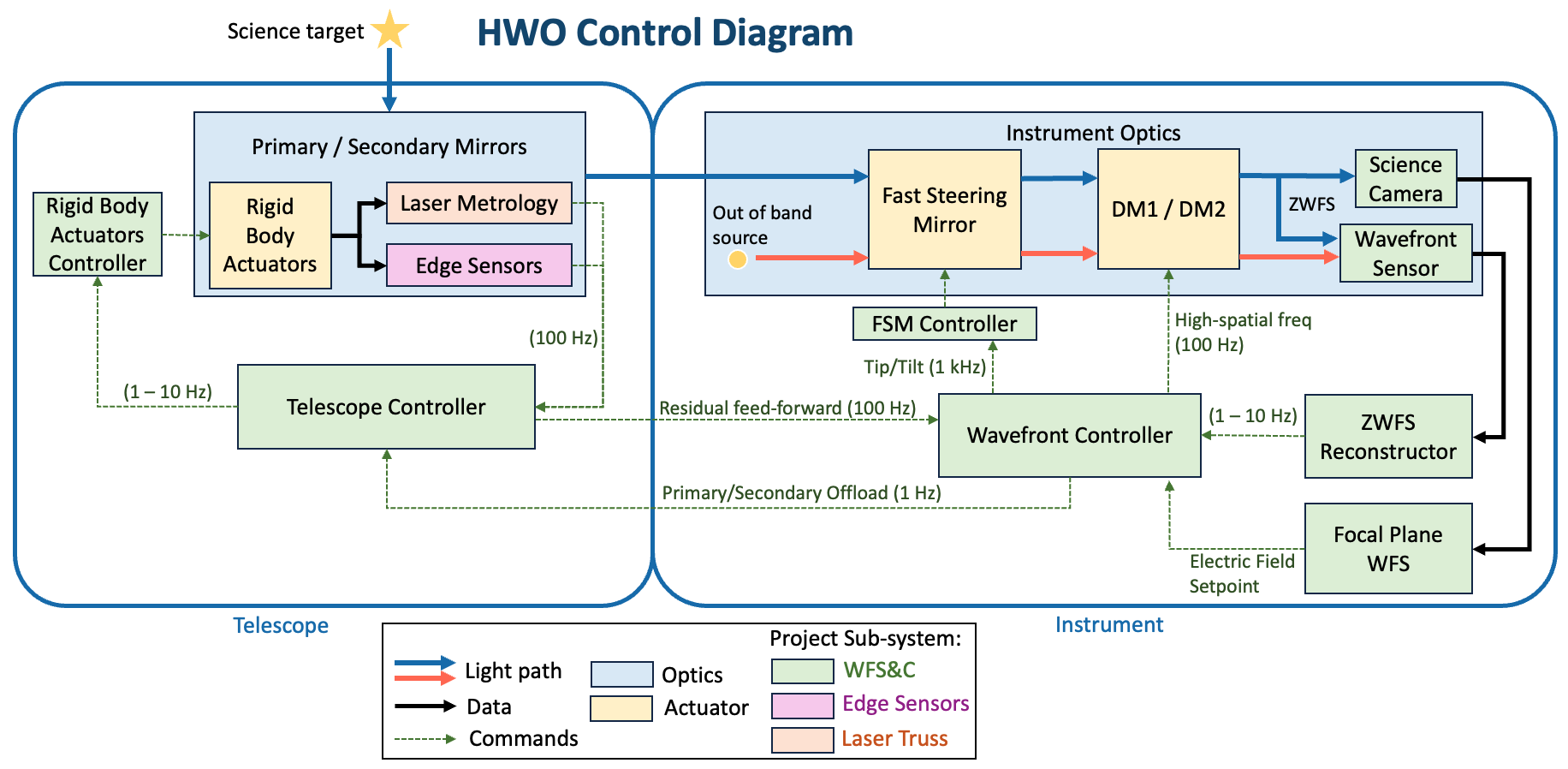} \\
\includegraphics[width=0.95\textwidth]{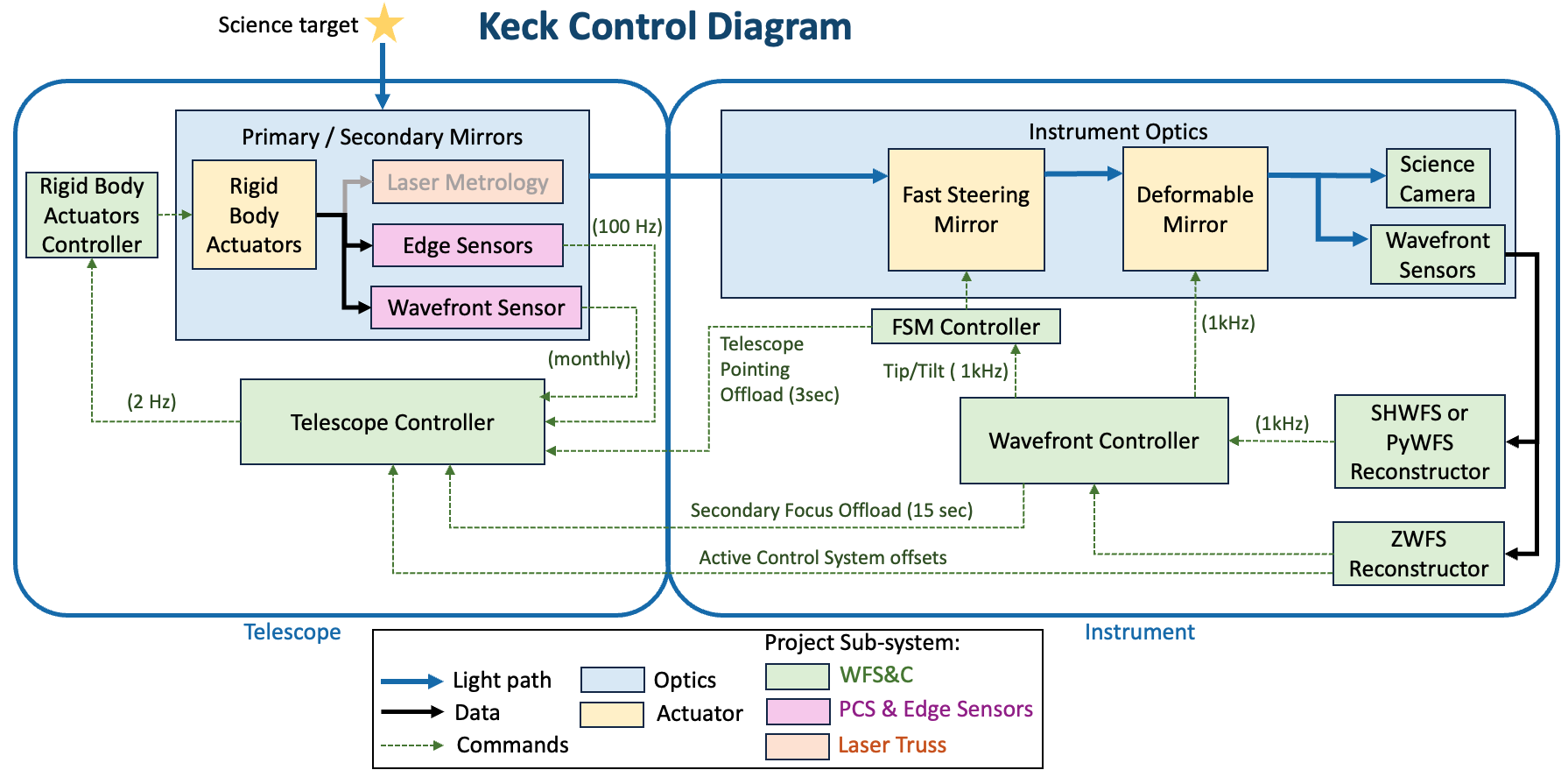}
\end{tabular}
\end{center}
\caption{\small \textit{Top:} HWO control architecture diagram (modified from \citealt{Feinberg23}). \textit{Bottom:} Keck control architecture (modified from \citealt{Chin22}). Highlighting the parallel architecture and subsystems between HWO and Keck. Although operating at different speeds and in a different wavefront error regime, Keck can be used to develop and validate nested wavefront sensing and control loop strategies. With the exception of the laser truss system, we can take advantage of existing Keck infrastructure to demonstrate the full system-level segment control architecture for HWO.
\label{fig:control_diagrams}
}
\end{figure*}

We now aim to take advantage of the full range of design similarities between the Keck and HWO architectures. Both observatories have a large segmented aperture primary mirror equipped with capacitive edge sensors for segment co-phasing and an active control system to maintain the phasing. Both observatories have a subsystem for WFS\&C with downstream high-contrast science instruments. Additionally, both architectures could potentially host a laser truss system. These hardware parallels enable similar control architectures, illustrated in Figure \ref{fig:control_diagrams}.

Given the complexity of these systems, the ULTRA team also identified the importance of considering the interactions between subsystems and the impacts on stability and requirements they have on each other. While Keck will not operate at picometer residuals, we plan to fully address concerns related to control authority, actuator offload, and stability. Moreover, even at Keck's nanometer-level phasing capabilities, successful comparisons of observed and predicted performances will validate, on a real operating observatory, the HWO error budget methodology and in particular its approach to nested loops operating at multiple timescales. 

\subsection{Goals}
We build on Keck's successful history of space-relevant technology demonstrations. For example, JWST's dispersed fringe sensing technology was experimentally validated on the Keck telescope \citep{Shi04}. We use the hardware parallels between Keck and HWO illustrated in Figure \ref{fig:control_diagrams} to address three goals:\\ 
\indent 1) Demonstrate closed-loop primary mirror control using a combination of ZWFS and edge sensor measurements in parallel with high contrast science data acquisition.\\
\indent 2) Identify barriers to sub-nanometer segment edge phasing between two Keck segments.\\
\indent 3) Complete a trade study of all sensors and control algorithms tested and translate these results into recommendations for HWO architecture and control scheme designs.

\section{Segment Phasing Approaches on Keck}

\subsection{Phasing Camera System and Capacitive Edge Sensors}

The Keck telescopes are aligned and phased using the Phasing Camera System (PCS; \citealt{Chanan89}). The phasing is maintained by the Active Control System (ACS; \citealt{Cohen94}), which adjusts the piston, tip, and tilt of each Keck segment. The ACS consists of capacitive displacement sensors which measure the relative height difference between two adjacent segment edges and send commands to the actuators to maintain the segments within predetermined sensor alignment readings. These desired sensor values are determined via a combination of values set by PCS, telescope Zenith angle, and temperature. The ACS sensors are sampled at 100Hz and the loop commands the actuators at a rate of 2Hz. The actuator step size is 4.18 nm. ACS segment positions are commanded via offsets in actuator space. These offsets are then converted to changes in the desired sensor readings and the desired sensor readings are updated. In closed loop the segments then move to achieve the desired motion.

PCS is a modified Shack-Hartmann camera mounted on the left bent cassegrain focus of the telescope. After a mirror segment has been exchanged for maintenance, PCS is used to measure segment figures, which are corrected via warping harness, as well as to align the secondary mirror and segments in tip/tilt and piston. PCS is also used approximately once a month to maintain the telescope alignment and align the secondary mirror and segments in tip/tilt and piston. PCS measures the segment pistons by measuring the phase error at 78 edge-sampled spots, each 120 mm in diameter \citep{Chanan98,Chanan00}. These 78 measurements are used to calculate the 36 segment pistons.

\subsection{Zernike Wavefront Sensor}

The Zernike wavefront sensor (ZWFS) was the baseline WFS for the HabEx and LUVOIR concept studies and is currently a key candidate wavefront sensor for HWO. Commonly used WFS technologies like the Shack-Hartmann and modulated Pyramid measure gradients in a wavefront and are therefore not equipped to accurately measure discontinuous phase aberrations caused by primary mirror segment misalignments. The ZWFS uses the phase-contrast technique, converting phase variations into intensity variations that can be imaged. The ZWFS consists of a focal-plane mask which offsets the phase of the core of the point spread function (PSF), causing it to interfere with the rest of the PSF light. The wavefront phase aberrations are thus encoded in the resulting intensity pattern. Due to this effect, the ZWFS is the most sensitive wavefront sensor \citep{Guyon05,Chambouleyron21} and can measure the discontinuous phase aberrations caused by a segmented mirror. The ZWFS will be mission tested in space for the first time on the Nancy Grace Roman Space Telescope \citep{Shi16,Riggs21},

We are using a vector-Zernike WFS (vZWFS), in which two different phase offsets are simultaneously imposed on two orthogonal light polarizations. This is enabled by using a mask made of metasurface optics (see \citealt{Wallace23} and \citealt{Wenger25} for more details). This design produces two pupil images, extending the otherwise very limited dynamic range of the ZWFS (as illustrated in Figure 4 of \citealt{Salama24}).

The Keck II telescope is equipped with an AO system running a visible Shack-Hartmann WFS (SHWFS) to sense atmospheric turbulence at kHz speeds. The vZWFS is installed downstream the Keck II AO system, in the Keck Planet Imager and Characterizer (KPIC; \citealt{Mawet16,Delorme21,Jovanovic25}) instrument, shown in Figure 2 of \cite{Salama24}. The NIRC2 instrument is used for high-contrast imaging with vortex and Lyot coronagraphs. The vZWFS can be used in parallel with NIRC2 science observations to sense primary segment cophasing errors while the SHWFS senses atmospheric turbulence. The segment piston corrections are applied by sending offsets to the ACS.

\section{Keck On-sky Testing}

\subsection{Wavefront Sensing \& Control} 

\subsubsection{Dual-Mirror Control}

Our existing demonstrations of measuring segment co-phasing errors with the ZWFS and correcting them with the primary mirror (illustrated in Figure \ref{fig:ZWFS_onsky} and detailed in \citealt{Salama24,Salama24b,Salama25}) assume that the AO system – which is upstream of the ZWFS – did not introduce any aberrations or correct for any co-phasing errors. While the AO system's SHWFS does not generally sense phase discontinuities, it can misinterpret the piston difference between two segments that happen to straddle a SHWFS sub-aperture as a local tip/tilt, which is then corrected by the DM. In \cite{Salama25}, we preliminarily addressed the impact of the AO system on the ZWFS measurements by applying known offsets to the segment pistons before running the ZWFS closed-loop segment piston control. Next, we aim to take full advantage of the vZWFS's sensitivity by disentangling segment co-phasing from other AO static or quasi-static aberrations so that co-phasing errors can be corrected with the primary mirror while other static aberrations are corrected with the DM. In this ``dual-mirror control" approach, the SHWFS will still be used to update the DM at kHz speeds, but the ZWFS is used to send additional offsets to the DM (and primary mirror) at slower speeds ($\sim$ minutes).

Because the ZWFS senses the wavefront error after it has been corrected by the DM, we will need to develop a control algorithm to determine which aberrations should be corrected with the primary mirror versus the DM. We will approach this problem by analyzing the full Keck AO telemetry to extract the co-phasing errors that are potentially being sensed by the SHWFS. We tested a potential method for disentangling co-phasing from downstream aberrations by running our vZWFS closed-loop tests at different Keck pupil rotation angles (briefly shown in \citealt{Salama25}), with the analysis still ongoing. We will develop and test our dual-mirror control algorithms on the SEAL testbed (see Section \S \ref{sec:SEAL}), before testing them on-sky on Keck. Once the dual mirror-control has been demonstrated to work on Keck and yields improved high-contrast science performance (e.g improved Strehl ratio and contrast), we will translate our findings into recommendations for HWO dual-mirror control schemes.

\subsubsection{Dual-Mirror, Dual-Sensor Control} 

The next step will be to incorporate PCS and edge sensor data into our dual-mirror control scheme. We aim to combine edge sensor data with ZWFS data to inform which errors originate from the primary mirror as opposed to AO residuals that should be corrected with the DM. The purpose is to determine how to best take advantage of the strengths of both the edge sensor data and the ZWFS data. We will first address this task using the SEAL testbed, where we can simulate the effects of PCS measurement errors by pistoning and tilting the segments of the Keck-geometry segmented DM. However, an important limitation of this approach is that the DM segments are flat while Keck (and HWO) segments are curved. After initial algorithm development on SEAL, we will therefore move onto observatory testing of the dual-mirror, dual-sensor approach, where the goal will be to improve the Strehl ratio and contrast compared with the dual-mirror control.

There is a fundamental difference between segment phasing with PCS, which measures segment edges, and the ZWFS, which measures the average piston error across the whole segment. While the Keck segments have no significant edge effects \citep{Troy18}, the segments do have on average 45~nm RMS of surface aberrations. Our goal is to investigate which solution (or combination thereof) provides the highest contrast science image post-AO correction – this is a key question for HWO. In \cite{Salama24b}, we ran both PCS and the ZWFS on the same night and found that the resulting phasing solutions were different. As part of the ZWFS effort, algorithms to phase the segments using the phase extracted around the edge of the segments versus the average segment piston will also be investigated.

\subsubsection{WFS\&C in Parallel with Science} 

Obtaining contemporaneous images on the science instruments will enable us to validate the WFS\&C strategies that yield and maintain the best PSF quality and highest contrast, rather than only measuring what yields the best phasing. The result will be the first full-scale, end-to-end demonstration of an example HWO segment control architecture.

\subsubsection{Santa cruz Extreme Adaptive optics Laboratory (SEAL)}
\label{sec:SEAL}

The SEAL testbed \citep{JensenClem21,JensenClem25} is designed to develop cutting-edge AO technologies for high-contrast imaging on ground-based segmented telescopes, such as Keck and future extremely large telescopes (ELTs). It contains three DMs: a 37-segmented mirror, mimicking the Keck aperture, and a woofer-tweeter system with a low-order 11-actuator DM and a high-order 648-actuator DM. The SEAL testbed is also equipped with a Spatial Light Modulator (SLM) used to simulate atmospheric turbulence and multiple wavefront sensors, including a SHWFS and a vZWFS \citep{Doelman19,Salama22,Chambouleyron24b}. 

We will use the SHWFS on SEAL to close the AO loop on the high-order DM and use the vZWFS to measure the AO residuals, disentangling aberrations originating from statics or quasi-statics in the AO system from segment co-phasing errors. Making use of similar architectures between Keck and SEAL, we can test vZWFS signal reconstruction, disentangling error sources, and control algorithms using the vZWFS downstream from the AO system. Finally, with the vortex coronagraph installed on SEAL \citep{Moreno24}, we can compare the impact of different reconstructors and control algorithms on the contrast.

\subsection{Towards Sub-Nanometer Two-Segment Edge Phasing Repeatability}

In early experiments, over a decade ago, the Keck PCS team demonstrated repeatable co-alignment of segment edges to 9~nm RMS, resulting in a piston error of 2~nm. 
As a reference, analysis of JWST WFS measurements shows a repeatability of 11~nm in 2024 (ignoring segment tilt events). HWO, however, will require significantly better performance. 

Here, we plan to investigate the phasing limitations of Keck's existing PCS technology and use this information to validate phasing simulations. We aim to re-implement the above experiment that demonstrated a repeatable piston error of 2~nm and investigate the barriers to pushing this value below 1~nm. Our first step will be to use existing simulations to understand the predicted on-sky performance for both the PCS and ZWFS phasing techniques. 
Further on-sky experiments will be performed to characterize the measurement repeatability of each phasing technique and compare each technique's absolute phase measurements. By combining these results with validated PCS and ZWFS simulations, we aim to understand the performance limitations of each technique in a realistic observatory environment.

\section{Analysis for HWO}

The final goal is to translate our results into HWO simulation inputs and architecture recommendations. 

\subsection{Trade Study of all sensing methods and control algorithms}

We will perform a trade study of the different sensing and phasing techniques from ZWFS and PCS/edge sensors. This trade study will apply diagnostic tools, such as noise rejection and residual errors due to fundamental limits, to all experimental and simulated data considered in this project (ZWFS results from SEAL and Keck, combined with the Keck results and validated simulations from the PCS and edge sensor tests). This will result in a comparison of the strengths and challenges of each sensing method and how they can be used together, in order to relax requirements on different subsystems, and to avoid exacerbating requirements on each other. 

\subsection{HWO architecture and control recommendations}

This project will represent the first time that complex control architectures with multiple sensor technologies and mirrors are fielded on a segmented telescope. Hence, the end goal of this project consists of using the results from these experiments at Keck to quantify key critical metrics that can then be translated to an HWO architecture, including (1) noise rejection functions, and (2) the ratio of measured residual errors to fundamental limits. The noise rejection function is the ratio of the closed loop and open loop wavefront residuals at a series of measurable timescales. It can be directly derived based on the proposed experiments. The second metric will inform how far the overall system operates from optimal performances. At Keck, wavefront estimation will be limited by noise in the wavefront estimate due to limited natural guide star photons (stellar magnitude plus uncorrected atmospheric turbulence) and noise in the edge sensors. Wavefront control will be limited by actuator resolution and stability, both for the primary actuators and DM. Control limits have been well characterized in the past at Keck. This project will quantify the estimation limits. Component-level limitations will also be folded in. We will predict closed loop residuals for each of the estimation/control architectures and compare them to observed residuals with Keck. To do so, we will use the formalism developed in the context of ultra-stable telescope studies \citep{Laginja20,Pogorelyuk21,Sahoo22,Coyle22}. In addition to achieving the first full system demonstration at the $\sim$nanometer level, we will go further by quantifying whether the performance limitations are actually driven by the specific Keck components and subsystems or if there still exist system level (e.g. control algorithms, architecture) performance gaps. \\

{\bf Acknowledgements.} This work is supported by NASA through the Astrophysics Research and Analysis (APRA) program (grant \#80NSSC25K7486).

\bibliography{author.bib}

\begin{thebibliography}{}
\parskip=0pt \itemsep=0pt \small \baselineskip=11pt
\expandafter\ifx\csname natexlab\endcsname\relax\def\natexlab#1{#1}\fi
\providecommand{\url}[1]{\href{#1}{#1}}
\providecommand{\dodoi}[1]{}
\providecommand{\doeprint}[1]{\href{http://ascl.net/#1}{#1}}
\providecommand{\doarXiv}[1]{\href{https://arxiv.org/abs/#1}{arXiv:#1}}

\bibitem[{{Beuzit} {et~al.}(2019){Beuzit}, {Vigan}, {Mouillet}, {Dohlen}, {Gratton}, {Boccaletti}, {Sauvage}, {Schmid}, {Langlois}, {Petit}, {Baruffolo}, {Feldt}, {Milli}, {Wahhaj}, {Abe}, {Anselmi}, {Antichi}, {Barette}, {Baudrand}, {Baudoz}, {Bazzon}, {Bernardi}, {Blanchard}, {Brast}, {Bruno}, {Buey}, {Carbillet}, {Carle}, {Cascone}, {Chapron}, {Charton}, {Chauvin}, {Claudi}, {Costille}, {De Caprio}, {de Boer}, {Delboulb{\'e}}, {Desidera}, {Dominik}, {Downing}, {Dupuis}, {Fabron}, {Fantinel}, {Farisato}, {Feautrier}, {Fedrigo}, {Fusco}, {Gigan}, {Ginski}, {Girard}, {Giro}, {Gisler}, {Gluck}, {Gry}, {Henning}, {Hubin}, {Hugot}, {Incorvaia}, {Jaquet}, {Kasper}, {Lagadec}, {Lagrange}, {Le Coroller}, {Le Mignant}, {Le Ruyet}, {Lessio}, {Lizon}, {Llored}, {Lundin}, {Madec}, {Magnard}, {Marteaud}, {Martinez}, {Maurel}, {M{\'e}nard}, {Mesa}, {M{\"o}ller-Nilsson}, {Moulin}, {Moutou}, {Orign{\'e}}, {Parisot}, {Pavlov}, {Perret}, {Pragt}, {Puget}, {Rabou}, {Ramos}, {Reess}, {Rigal}, {Rochat}, {Roelfsema}, {Rousset},
  {Roux}, {Saisse}, {Salasnich}, {Santambrogio}, {Scuderi}, {Segransan}, {Sevin}, {Siebenmorgen}, {Soenke}, {Stadler}, {Suarez}, {Tiph{\`e}ne}, {Turatto}, {Udry}, {Vakili}, {Waters}, {Weber}, {Wildi}, {Zins}, \& {Zurlo}}]{Beuzit19}
{Beuzit}, J.~L., {Vigan}, A., {Mouillet}, D., {et~al.} 2019, \href{http://doi.org/10.1051/0004-6361/201935251}{\color{blue}\aap}, \href{https://ui.adsabs.harvard.edu/abs/2019A&A...631A.155B}{\color{blue}631}, A155

\bibitem[{{Chambouleyron} {et~al.}(2021){Chambouleyron}, {Fauvarque}, {Sauvage}, {Dohlen}, {Levraud}, {Vigan}, {N'Diaye}, {Neichel}, \& {Fusco}}]{Chambouleyron21}
{Chambouleyron}, V., {Fauvarque}, O., {Sauvage}, J.~F., {et~al.} 2021, \href{http://doi.org/10.1051/0004-6361/202140870}{\color{blue}\aap}, \href{https://ui.adsabs.harvard.edu/abs/2021A&A...650L...8C}{\color{blue}650}, L8

\bibitem[{{Chambouleyron} {et~al.}(2024){Chambouleyron}, {Ciss{\'e}}, {Salama}, {Haffert}, {D{\'e}o}, {Guthery}, {Wallace}, {Dillon}, {Jensen-Clem}, {Hinz}, \& {Macintosh}}]{Chambouleyron24b}
{Chambouleyron}, V., {Ciss{\'e}}, M., {Salama}, M., {et~al.} 2024, in Society of Photo-Optical Instrumentation Engineers (SPIE) Conference Series, Vol. 13097, \href{http://doi.org/10.1117/12.3020670}{\color{blue}Adaptive Optics Systems IX}, ed. K.~J. {Jackson}, D.~{Schmidt}, \& E.~{Vernet}, 130971N

\bibitem[{{Chanan} {et~al.}(2000){Chanan}, {Ohara}, \& {Troy}}]{Chanan00}
{Chanan}, G., {Ohara}, C., \& {Troy}, M. 2000, \href{http://doi.org/10.1364/AO.39.004706}{\color{blue}\ao}, \href{https://ui.adsabs.harvard.edu/abs/2000ApOpt..39.4706C}{\color{blue}39}, 4706

\bibitem[{{Chanan} {et~al.}(1998){Chanan}, {Troy}, {Dekens}, {Michaels}, {Nelson}, {Mast}, \& {Kirkman}}]{Chanan98}
{Chanan}, G., {Troy}, M., {Dekens}, F., {et~al.} 1998, \href{http://doi.org/10.1364/AO.37.000140}{\color{blue}\ao}, \href{https://ui.adsabs.harvard.edu/abs/1998ApOpt..37..140C}{\color{blue}37}, 140

\bibitem[{{Chanan}(1989)}]{Chanan89}
{Chanan}, G.~A. 1989, in Society of Photo-Optical Instrumentation Engineers (SPIE) Conference Series, Vol. 1036, \href{http://doi.org/10.1117/12.950971}{\color{blue}Precision Instrument Design}, ed. T.~C. {Bristow} \& A.~E. {Hatheway}, 59

\bibitem[{{Chin} {et~al.}(2022){Chin}, {Cetre}, {Wizinowich}, {Ragland}, {Lilley}, {Wetherell}, {Surendran}, {Correia}, {Marin}, {Biasi}, {Pataunar}, {Pescoller}, {Glazebrook}, {Jameson}, {Gauvin}, {Rigaut}, {Gratadour}, \& {Bernard}}]{Chin22}
{Chin}, J. C.~Y., {Cetre}, S., {Wizinowich}, P., {et~al.} 2022, in Society of Photo-Optical Instrumentation Engineers (SPIE) Conference Series, Vol. 12185, \href{http://doi.org/10.1117/12.2629614}{\color{blue}Adaptive Optics Systems VIII}, ed. L.~{Schreiber}, D.~{Schmidt}, \& E.~{Vernet}, 121850V

\bibitem[{{Cohen} {et~al.}(1994){Cohen}, {Mast}, \& {Nelson}}]{Cohen94}
{Cohen}, R.~W., {Mast}, T.~S., \& {Nelson}, J.~E. 1994, in Society of Photo-Optical Instrumentation Engineers (SPIE) Conference Series, Vol. 2199, \href{http://doi.org/10.1117/12.176173}{\color{blue}Advanced Technology Optical Telescopes V}, ed. L.~M. {Stepp}, 105--116

\bibitem[{{Coyle} {et~al.}(2019){Coyle}, {Knight}, {Barto}, {Allard}, {Lipscy}, {East}, {Wells}, {Havey}, {Sullivan}, {Allen}, {Arenberg}, {Lawton}, {Patton}, {Hellekson}, {Van Otten}, {Bluth}, {Nielsen}, {Pueyo}, \& {Soummer}}]{ULTRA}
{Coyle}, L., {Knight}, S., {Barto}, A., {et~al.} 2019, in Bulletin of the American Astronomical Society, Vol.~51, 80

\bibitem[{{Coyle} {et~al.}(2022){Coyle}, {Knight}, {Pueyo}, {East}, {Hellekson}, {Bluth}, {Park}, {Tucker}, {Hicks}, {Cromey}, {Sahoo}, {Soummer}, {Brennan}, {Lawton}, {Arenberg}, \& {Eisenhower}}]{Coyle22}
{Coyle}, L.~E., {Knight}, J.~S., {Pueyo}, L., {et~al.} 2022, in Society of Photo-Optical Instrumentation Engineers (SPIE) Conference Series, Vol. 12180, \href{http://doi.org/10.1117/12.2627057}{\color{blue}Space Telescopes and Instrumentation 2022: Optical, Infrared, and Millimeter Wave}, ed. L.~E. {Coyle}, S.~{Matsuura}, \& M.~D. {Perrin}, 121802K

\bibitem[{{Cromey} {et~al.}(2022){Cromey}, {Hicks}, {Shugrue}, {Ho}, {Owen}, {Grossman}, {Fan}, {Walters}, {Knight}, \& {Coyle}}]{Cromey22}
{Cromey}, B., {Hicks}, B., {Shugrue}, J., {et~al.} 2022, in Society of Photo-Optical Instrumentation Engineers (SPIE) Conference Series, Vol. 12180, \href{http://doi.org/10.1117/12.2628183}{\color{blue}Space Telescopes and Instrumentation 2022: Optical, Infrared, and Millimeter Wave}, ed. L.~E. {Coyle}, S.~{Matsuura}, \& M.~D. {Perrin}, 121805Y

\bibitem[{{Delorme} {et~al.}(2021){Delorme}, {Jovanovic}, {Echeverri}, {Mawet}, {Kent Wallace}, {Bartos}, {Cetre}, {Wizinowich}, {Ragland}, {Lilley}, {Wetherell}, {Doppmann}, {Wang}, {Morris}, {Ruffio}, {Martin}, {Fitzgerald}, {Ruane}, {Schofield}, {Suominen}, {Calvin}, {Wang}, {Magnone}, {Johnson}, {Sohn}, {L{\'o}pez}, {Bond}, {Pezzato}, {Sayson}, {Chun}, \& {Skemer}}]{Delorme21}
{Delorme}, J.-R., {Jovanovic}, N., {Echeverri}, D., {et~al.} 2021, \href{http://doi.org/10.1117/1.JATIS.7.3.035006}{\color{blue}Journal of Astronomical Telescopes, Instruments, and Systems}, \href{https://ui.adsabs.harvard.edu/abs/2021JATIS...7c5006D}{\color{blue}7}, 035006

\bibitem[{{Doelman} {et~al.}(2019){Doelman}, {Fagginger Auer}, {Escuti}, \& {Snik}}]{Doelman19}
{Doelman}, D.~S., {Fagginger Auer}, F., {Escuti}, M.~J., {et~al.} 2019, \href{http://doi.org/10.1364/OL.44.000017}{\color{blue}Optics Letters}, \href{https://ui.adsabs.harvard.edu/abs/2019OptL...44...17D}{\color{blue}44}, 17

\bibitem[{{Feinberg} {et~al.}(2023){Feinberg}, {Coyle}, {Redding}, {Sitarski}, {Tesch}, {Menzel}, {Park}, {Bluth}, \& {van Campen}}]{Feinberg23}
{Feinberg}, L., {Coyle}, L., {Redding}, D., {et~al.} 2023, in {Ultrastable Observatory Roadmap Team (USORT) Status}.
\newblock \url{https://ntrs.nasa.gov/citations/20230011036}

\bibitem[{{Guyon}(2005)}]{Guyon05}
{Guyon}, O. 2005, \href{http://doi.org/10.1086/431209}{\color{blue}\apj}, \href{https://ui.adsabs.harvard.edu/abs/2005ApJ...629..592G}{\color{blue}629}, 592

\bibitem[{Jensen-Clem {et~al.}(2021)Jensen-Clem, Dillon, Gerard, van Kooten, Fowler, Kupke, Cetre, Sanchez, Hinz, Laguna, Doelman, \& Snik}]{JensenClem21}
Jensen-Clem, R., Dillon, D., Gerard, B., {et~al.} 2021, in \href{http://doi.org/10.1117/12.2594676}{\color{blue}Techniques and Instrumentation for Detection of Exoplanets X}, ed. S.~B. Shaklan \& G.~J. Ruane, Vol. 11823, International Society for Optics and Photonics (SPIE), 118231D

\bibitem[{{Jensen-Clem} {et~al.}(2025){Jensen-Clem}, {Chambouleyron}, {Javier}, {Dillon}, {Por}, {Calvin}, {Cetre}, {Amezcua Correa}, {Crowe}, {Diaz}, {Dobias}, {Doelman}, {Eikenberry}, {Fowler}, {Gerard}, {Hinz}, {Kupke}, {Moreno}, {Nguyen}, {Salama}, {Sengupta}, {Skaf}, \& {Snik}}]{JensenClem25}
{Jensen-Clem}, R., {Chambouleyron}, V., {Javier}, P., {et~al.} 2025

\bibitem[{{Jovanovic} {et~al.}(2025){Jovanovic}, {Echeverri}, {Delorme}, {Finnerty}, {Schofield}, {Wang}, {Xin}, {Xuan}, {Wallace}, {Mawet}, {Sanghi}, {Baker}, {Bartos}, {Bond}, {Calvin}, {Cetre}, {Doppmann}, {Fitzgerald}, {Fucik}, {Gao}, {Ge}, {Guthery}, {Horstman}, {Hsu}, {Liberman}, {Leifer}, {Lilley}, {Lopez}, {Marin}, {Martin}, {Mennesson}, {Morris}, {Nash}, {Pezzato}, {Porter}, {Roberts}, {Ruane}, {Ruffio}, {Sappey}, {Serabyn}, {Shen}, {Skemer}, {Wang}, {Wetherell}, {Wizinowich}, {Salama}, {Chambouleyron}, {Jensen-Clem}, \& {Beichman}}]{Jovanovic25}
{Jovanovic}, N., {Echeverri}, D., {Delorme}, J.-R., {et~al.} 2025, \href{http://doi.org/10.1117/1.JATIS.11.1.015005}{\color{blue}Journal of Astronomical Telescopes, Instruments, and Systems}, \href{https://ui.adsabs.harvard.edu/abs/2025JATIS..11a5005J}{\color{blue}11}, 015005

\bibitem[{{Laginja} {et~al.}(2020){Laginja}, {Soummer}, {Mugnier}, {Pueyo}, {Sauvage}, {Leboulleux}, {Coyle}, {Knight}, {Perrin}, {Will}, {Noss}, {Brooks}, \& {Fowler}}]{Laginja20}
{Laginja}, I., {Soummer}, R., {Mugnier}, L.~M., {et~al.} 2020, in Society of Photo-Optical Instrumentation Engineers (SPIE) Conference Series, Vol. 11443, \href{http://doi.org/10.1117/12.2560113}{\color{blue}Space Telescopes and Instrumentation 2020: Optical, Infrared, and Millimeter Wave}, ed. M.~{Lystrup} \& M.~D. {Perrin}, 114433J

\bibitem[{{Macintosh} {et~al.}(2018){Macintosh}, {Chilcote}, {Bailey}, {de Rosa}, {Nielsen}, {Norton}, {Poyneer}, {Wang}, {Ruffio}, {Graham}, {Marois}, {Savransky}, \& {Veran}}]{Macintosh18}
{Macintosh}, B., {Chilcote}, J.~K., {Bailey}, V.~P., {et~al.} 2018, in Society of Photo-Optical Instrumentation Engineers (SPIE) Conference Series, Vol. 10703, \href{http://doi.org/10.1117/12.2314253}{\color{blue}Adaptive Optics Systems VI}, ed. L.~M. {Close}, L.~{Schreiber}, \& D.~{Schmidt}, 107030K

\bibitem[{{Mawet} {et~al.}(2016){Mawet}, {Wizinowich}, {Dekany}, {Chun}, {Hall}, {Cetre}, {Guyon}, {Wallace}, {Bowler}, {Liu}, {Ruane}, {Serabyn}, {Bartos}, {Wang}, {Vasisht}, {Fitzgerald}, {Skemer}, {Ireland}, {Fucik}, {Fortney}, {Crossfield}, {Hu}, \& {Benneke}}]{Mawet16}
{Mawet}, D., {Wizinowich}, P., {Dekany}, R., {et~al.} 2016, in Society of Photo-Optical Instrumentation Engineers (SPIE) Conference Series, Vol. 9909, \href{http://doi.org/10.1117/12.2233658}{\color{blue}Adaptive Optics Systems V}, ed. E.~{Marchetti}, L.~M. {Close}, \& J.-P. {V{\'e}ran}, 99090D

\bibitem[{{Moreno} {et~al.}(2024){Moreno}, {Chambouleyron}, {Jensen-Clem}, {Dillon}, {Hinz}, \& {Macintosh}}]{Moreno24}
{Moreno}, A., {Chambouleyron}, V., {Jensen-Clem}, R.~M., {et~al.} 2024, in Society of Photo-Optical Instrumentation Engineers (SPIE) Conference Series, Vol. 13097, \href{http://doi.org/10.1117/12.3020445}{\color{blue}Adaptive Optics Systems IX}, ed. K.~J. {Jackson}, D.~{Schmidt}, \& E.~{Vernet}, 1309746

\bibitem[{{Nemati} {et~al.}(2020){Nemati}, {Stahl}, {Stahl}, {Ruane}, \& {Sheldon}}]{Nemati20}
{Nemati}, B., {Stahl}, H.~P., {Stahl}, M.~T., {et~al.} 2020, \href{http://doi.org/10.1117/1.JATIS.6.3.039002}{\color{blue}Journal of Astronomical Telescopes, Instruments, and Systems}, \href{https://ui.adsabs.harvard.edu/abs/2020JATIS...6c9002N}{\color{blue}6}, 039002

\bibitem[{{Pogorelyuk} {et~al.}(2021){Pogorelyuk}, {Pueyo}, {Males}, {Cahoy}, \& {Kasdin}}]{Pogorelyuk21}
{Pogorelyuk}, L., {Pueyo}, L., {Males}, J.~R., {et~al.} 2021, \href{http://doi.org/10.3847/1538-4365/ac126d}{\color{blue}\apjs}, \href{https://ui.adsabs.harvard.edu/abs/2021ApJS..256...39P}{\color{blue}256}, 39

\bibitem[{{Potier} {et~al.}(2021){Potier}, {Ruane}, {Chen}, {Chopra}, {Dewell}, {Juanola Parramon}, {Nordt}, {Pueyo}, {Redding}, {Eldorado Riggs}, \& {Sirbu}}]{Potier21}
{Potier}, A., {Ruane}, G., {Chen}, P., {et~al.} 2021, in Society of Photo-Optical Instrumentation Engineers (SPIE) Conference Series, Vol. 11823, \href{http://doi.org/10.1117/12.2595116}{\color{blue}Techniques and Instrumentation for Detection of Exoplanets X}, ed. S.~B. {Shaklan} \& G.~J. {Ruane}, 118231L

\bibitem[{{Potier} {et~al.}(2022){Potier}, {Ruane}, {Stark}, {Chen}, {Chopra}, {Dewell}, {Juanola-Parramon}, {Nordt}, {Pueyo}, {Redding}, {Eldorado Riggs}, \& {Sirbu}}]{Potier22}
{Potier}, A., {Ruane}, G., {Stark}, C., {et~al.} 2022, \href{http://doi.org/10.1117/1.JATIS.8.3.035002}{\color{blue}Journal of Astronomical Telescopes, Instruments, and Systems}, \href{https://ui.adsabs.harvard.edu/abs/2022JATIS...8c5002P}{\color{blue}8}, 035002

\bibitem[{{Pueyo} {et~al.}(2022){Pueyo}, {Juanola-Parramon}, {Tumlinson}, {Soummer}, {Laginja}, {Hammel}, \& {Mountain}}]{Pueyo22}
{Pueyo}, L., {Juanola-Parramon}, R., {Tumlinson}, J., {et~al.} 2022, \href{http://doi.org/10.1117/1.JATIS.8.4.049002}{\color{blue}Journal of Astronomical Telescopes, Instruments, and Systems}, \href{https://ui.adsabs.harvard.edu/abs/2022JATIS...8d9002P}{\color{blue}8}, 049002

\bibitem[{{Riggs} {et~al.}(2021){Riggs}, {Bailey}, {Moody}, {Sidick}, {Balasubramanian}, {Moore}, {Wilson}, {Ruane}, {Sirbu}, {Gersh-Range}, {Trauger}, {Mennesson}, {Siegler}, {Bendek}, {Groff}, {Zimmerman}, {Debes}, {Basinger}, \& {Kasdin}}]{Riggs21}
{Riggs}, A.~J.~E., {Bailey}, V., {Moody}, D.~C., {et~al.} 2021, in Society of Photo-Optical Instrumentation Engineers (SPIE) Conference Series, Vol. 11823, \href{http://doi.org/10.1117/12.2598599}{\color{blue}Techniques and Instrumentation for Detection of Exoplanets X}, ed. S.~B. {Shaklan} \& G.~J. {Ruane}, 118231Y

\bibitem[{{Sahoo} {et~al.}(2022){Sahoo}, {Laginja}, {Pueyo}, {Soummer}, {Coyle}, {Knight}, \& {East}}]{Sahoo22}
{Sahoo}, A., {Laginja}, I., {Pueyo}, L., {et~al.} 2022, in Society of Photo-Optical Instrumentation Engineers (SPIE) Conference Series, Vol. 12180, \href{http://doi.org/10.1117/12.2630592}{\color{blue}Space Telescopes and Instrumentation 2022: Optical, Infrared, and Millimeter Wave}, ed. L.~E. {Coyle}, S.~{Matsuura}, \& M.~D. {Perrin}, 121805V

\bibitem[{Salama {et~al.}(2022)Salama, Jensen-Clem, van Kooten, Dillon, Gerard, Fowler, Cetre, Snik, \& Doelman}]{Salama22}
Salama, M., Jensen-Clem, R., van Kooten, M., {et~al.} 2022, in \href{http://doi.org/10.1117/12.2630026}{\color{blue}Adaptive Optics Systems VIII}, ed. L.~Schreiber, D.~Schmidt, \& E.~Vernet, Vol. 12185, International Society for Optics and Photonics (SPIE), 121858M

\bibitem[{{Salama} {et~al.}(2024{\natexlab{a}}){Salama}, {Guthery}, {Chambouleyron}, {Jensen-Clem}, {Wallace}, {Delorme}, {Troy}, {Wenger}, {Echeverri}, {Finnerty}, {Jovanovic}, {Liberman}, {L{\'o}pez}, {Mawet}, {Morris}, {van Kooten}, {Wang}, {Wizinowich}, {Xin}, \& {Xuan}}]{Salama24}
{Salama}, M., {Guthery}, C., {Chambouleyron}, V., {et~al.} 2024{\natexlab{a}}, \href{http://doi.org/10.3847/1538-4357/ad3b99}{\color{blue}The Astrophysical Journal (ApJ)}, \href{https://ui.adsabs.harvard.edu/abs/2024ApJ...967..171S}{\color{blue}967}, 171

\bibitem[{{Salama} {et~al.}(2024{\natexlab{b}}){Salama}, {Guthery}, {Chambouleyron}, {Jensen-Clem}, {Wallace}, {Troy}, {Delorme}, {Dillon}, {Echeverri}, {Xin}, {Xuan}, {Jovanovic}, {Mawet}, {Wizinowich}, \& {Bowens-Rubin}}]{Salama24b}
{Salama}, M., {Guthery}, C.~E., {Chambouleyron}, V., {et~al.} 2024{\natexlab{b}}, in Society of Photo-Optical Instrumentation Engineers (SPIE) Conference Series, Vol. 13097, \href{http://doi.org/10.1117/12.3019373}{\color{blue}Adaptive Optics Systems IX}, ed. K.~J. {Jackson}, D.~{Schmidt}, \& E.~{Vernet}, 130971P

\bibitem[{Salama {et~al.}(2025)Salama, Ciss{\'e}, Chambouleyron, Guthery, Jensen-Clem, Wallace, Troy, Pueyo, Bouchez, Echeverri, Xin, Xuan, Jovanovic, \& Mawet}]{Salama25}
Salama, M., Ciss{\'e}, M., Chambouleyron, V., {et~al.} 2025, in \href{http://doi.org/10.1117/12.3064529}{\color{blue}Techniques and Instrumentation for Detection of Exoplanets XII}, ed. G.~J. Ruane \& M.~A. Millar-Blanchaer, Vol. 13627, International Society for Optics and Photonics (SPIE), 136270G

\bibitem[{{Seo} {et~al.}(2019){Seo}, {Patterson}, {Balasubramanian}, {Crill}, {Chui}, {Echeverri}, {Kern}, {Marx}, {Moody}, {Mejia Prada}, {Ruane}, {Shi}, {Shaw}, {Siegler}, {Tang}, {Trauger}, {Wilson}, \& {Zimmer}}]{Seo19}
{Seo}, B.-J., {Patterson}, K., {Balasubramanian}, K., {et~al.} 2019, in Society of Photo-Optical Instrumentation Engineers (SPIE) Conference Series, Vol. 11117, \href{http://doi.org/10.1117/12.2530033}{\color{blue}Society of Photo-Optical Instrumentation Engineers (SPIE) Conference Series}, 111171V

\bibitem[{{Shi} {et~al.}(2004){Shi}, {Chanan}, {Ohara}, {Troy}, \& {Redding}}]{Shi04}
{Shi}, F., {Chanan}, G., {Ohara}, C., {et~al.} 2004, \href{http://doi.org/10.1364/AO.43.004474}{\color{blue}\ao}, \href{https://ui.adsabs.harvard.edu/abs/2004ApOpt..43.4474S}{\color{blue}43}, 4474

\bibitem[{{Shi} {et~al.}(2016){Shi}, {Balasubramanian}, {Hein}, {Lam}, {Moore}, {Moore}, {Patterson}, {Poberezhskiy}, {Shields}, {Sidick}, {Tang}, {Truong}, {Wallace}, {Wang}, \& {Wilson}}]{Shi16}
{Shi}, F., {Balasubramanian}, K., {Hein}, R., {et~al.} 2016, \href{http://doi.org/10.1117/1.JATIS.2.1.011021}{\color{blue}Journal of Astronomical Telescopes, Instruments, and Systems}, \href{https://ui.adsabs.harvard.edu/abs/2016JATIS...2a1021S}{\color{blue}2}, 011021

\bibitem[{{Soummer} {et~al.}(2024){Soummer}, {Pourcelot}, {Por}, {Steiger}, {Laginja}, {Buralli}, {Redmond}, {Pueyo}, {Perrin}, {Ferrari}, {Fowler}, {Hagopian}, {N'Diaye}, {Nguyen}, {Nickson}, {Petrone}, {Sahoo}, {Sivaramakrishnan}, \& {Will}}]{Soummer24}
{Soummer}, R., {Pourcelot}, R., {Por}, E.~H., {et~al.} 2024, in Society of Photo-Optical Instrumentation Engineers (SPIE) Conference Series, Vol. 13092, \href{http://doi.org/10.1117/12.3018037}{\color{blue}Space Telescopes and Instrumentation 2024: Optical, Infrared, and Millimeter Wave}, ed. L.~E. {Coyle}, S.~{Matsuura}, \& M.~D. {Perrin}, 130921Z

\bibitem[{{Trauger} \& {Traub}(2007)}]{Trauger07}
{Trauger}, J.~T., \& {Traub}, W.~A. 2007, \href{http://doi.org/10.1038/nature05729}{\color{blue}\nat}, \href{https://ui.adsabs.harvard.edu/abs/2007Natur.446..771T}{\color{blue}446}, 771

\bibitem[{{Troy} {et~al.}(2018){Troy}, {Chanan}, {Colavita}, \& {Martinek}}]{Troy18}
{Troy}, M., {Chanan}, G., {Colavita}, M., {et~al.} 2018, in Society of Photo-Optical Instrumentation Engineers (SPIE) Conference Series, Vol. 10700, \href{http://doi.org/10.1117/12.2314568}{\color{blue}Ground-based and Airborne Telescopes VII}, ed. H.~K. {Marshall} \& J.~{Spyromilio}, 107000M

\bibitem[{{Wallace} {et~al.}(2023){Wallace}, {Wenger}, {Jewell}, {Salama}, {Chambouleyron}, {van Kooten}, {Guthery}, {Ragland}, {Delorme}, {Jensen-Clem}, {Wizinowich}, {Mawet}, \& {Jovanovic}}]{Wallace23}
{Wallace}, J., {Wenger}, T., {Jewell}, J., {et~al.} 2023, in Seventh International Conference on Adaptive Optics for Extremely Large Telescopes (AO4ELT), 9.
\newblock \url{https://ao4elt7.sciencesconf.org/503383}

\bibitem[{{Wenger} \& {Wallace}(2025)}]{Wenger25}
{Wenger}, T., \& {Wallace}, J.~K. 2025, \href{http://doi.org/10.1364/OL.544385}{\color{blue}Optics Letters}, \href{https://ui.adsabs.harvard.edu/abs/2025OptL...50..726W}{\color{blue}50}, 726

\end{thebibliography}

\end{document}